# Inertial and viscous flywheel sensing of nanoparticles

Georgios Katsikis,[a,b] Jesse F. Collis,[a,c] Scott M. Knudsen,[b] Vincent Agache,[b,d] John E. Sader,[c,e] Scott R. Manalis[b,f,g,h]

Rotational dynamics often challenge physical intuition while enabling unique realizations, from the rotor of a gyroscope that maintains its orientation regardless of the outer gimbals,[1] to a tennis racket that rotates around its handle when tossed face-up in the air.[2] In the context of inertial mass sensing, which can measure mass with atomic precision,[3] rotational dynamics are normally considered a complication hindering measurement interpretation. Here, we exploit the rotational dynamics of a microfluidic device to develop a new modality in inertial resonant sensing.[4–7] Combining theory with experiments, we show that this modality normally measures the volume of the particle while being insensitive to its density. Paradoxically, particle density only emerges when fluid viscosity becomes dominant over inertia. We explain this paradox via a **viscosity-driven, hydrodynamic coupling between the fluid and the particle that activates the rotational inertia of the particle, converting it into a 'viscous flywheel'**. This modality now enables the simultaneous measurement of particle volume and mass in fluid, using a single, high-throughput measurement.

Drawing from model paradigms in classical physics such as mass-spring systems to everyday objects such as guitar strings,[8] tuning forks[9] and bridge structures,[10] the natural or resonant frequency of a given object is intuitively associated with its mass and stiffness. Generally, the heavier and softer the object, the lower its natural frequency. In inertial mass sensing, thin plates or long cantilevers, either hollow or rigid, are driven to oscillate at their resonant frequency.[4,5,7,11,12] When a particle traverses inside or lands upon the surface of the sensor,[6,13] provided a local displacement exists, the particle changes the resonant frequency of the sensor in proportion to its mass.[4–7,11,14] Within this standard framework, rotational dynamics of the sensor are either ignored, or considered to be an erroneous or complicating factor in the measurement.[15]

Here, we exploit rotational dynamics in inertial sensing to enable a new measurement modality. We utilize the resonant frequency change induced when a rigid particle is suspended in fluid between microchannel walls that exhibit oscillatory rotation (Fig. 1a). We experimentally realize this motion using cantilevers in the form of Suspended Micro- and Nanochannel Resonators (SMRs/SNRs).[16,17] The cantilevers vibrate in their second resonant flexural mode, where local rotation, without displacement, occurs at the vibrational nodes.

Actuating the cantilever with no particle present at a vibrational node generates a flow within the microchannel, termed the "base flow".[18] This flow depends on the viscous penetration depth, $\delta$, relative to the flow channel height, $H$, and is quantified by the oscillatory Reynolds number, $\beta = (H/\delta)^2$; which also specifies the ratio of inertial-to-viscous forces in the fluid. The viscous penetration depth, $\delta = \sqrt{\mu/(\rho\omega)}$, is the distance from a solid boundary—here, the rotating walls—into the fluid where viscous effects are important. The fluid has shear viscosity, $\mu$, and mass density, $\rho$; $\omega = 2\pi f$ is the angular frequency of the wall motion, which here is a resonant frequency of the cantilever.

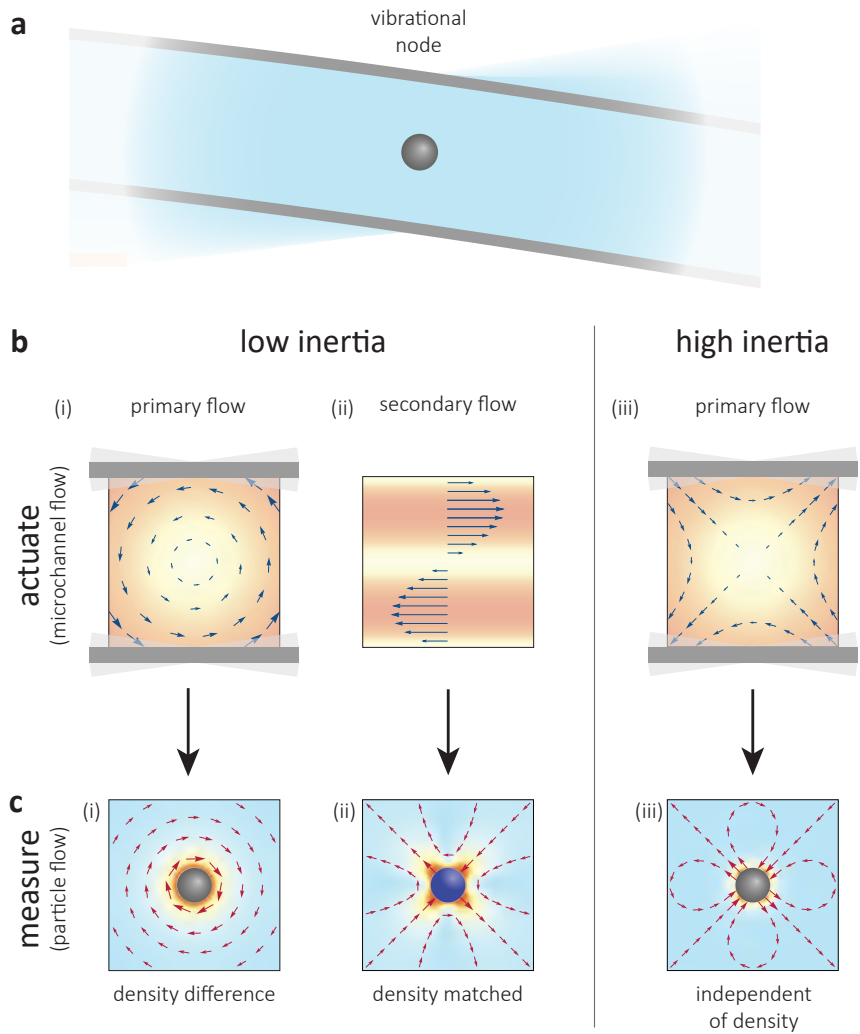

**Figure 1 | Concept of local rotation in inertial sensing. a,** Schematic of a rigid particle suspended at a vibrational node of a rotating microchannel. **b,** Actuating the walls with oscillatory rotation generates a flow within the microchannel or "base flow". (i) At low inertia, i.e., when viscous effects dominate, the base flow is primarily a rigid-body rotation. (ii) This primary flow generates a secondary, non-linear, shear flow due to the effects of small but finite inertia. (iii) At high inertia, i.e., where viscous effects are negligible, the base flow is primarily extensional. **c,** Suspending a particle in the base flow generates a "disturbance flow", which produces a measurable signal, Eq. (1). (i) At low inertia, a particle with a different density to the fluid rotates relative to the primary flow, generating a rotational disturbance flow, shown here for a negatively buoyant (heavier than fluid) particle. For a positively buoyant particle the direction of the disturbance flow field is reversed. (ii) A particle with its density matched to the fluid does not rotate relative to the primary flow, or to the rotational component of the secondary flow. It only reacts to the extensional component of the secondary flow (Fig. S1b-iii,iv), generating a quadrupole disturbance flow. (iii) At high inertia, the particle reacts to the extensional base flow, generating a quadrupole disturbance flow. This is in a similar manner to (ii), but independent of particle density.

[a] These authors contributed equally to the work.
[b] Koch Institute for Integrative Cancer Research, Massachusetts Institute of Technology, Cambridge, MA 02139 USA.
[c] ARC Centre of Excellence in Exciton Science, School of Mathematics and Statistics, The University of Melbourne, VIC 3010, Australia.
[d] Université Grenoble Alpes, CEA, LETI, 38000, Grenoble France.
[e] Corresponding author: jsader@unimelb.edu.au.
[f] Department of Biological Engineering, Massachusetts Institute of Technology, Cambridge, MA, 02139 USA.
[g] Department of Mechanical Engineering, Massachusetts Institute of Technology, Cambridge, MA 02139 USA.
[h] Corresponding author: srm@mit.edu.

To gain insight into the base flow, we explore the limits of low ($\beta \ll 1$) and high ($\beta \gg 1$) fluid inertia. For low inertia ($\beta \ll 1$), viscous effects dominate throughout the channel; the near-total absence of inertia in the fluid mostly occludes motion relative to the walls. Thus, the rotating walls generate a base flow that is primarily a rigid-body rotation (Fig. 1b-i). A finite, but small, amount of inertia in the fluid causes the fluid to slightly lag this primary base flow, producing a secondary shear flow that exhibits a nonlinear profile across the height of the channel (Fig. 1b-ii). This secondary flow can be decomposed into its fundamental rotational and extensional components (Fig. S1a, ii-v). The superposition of the primary and secondary base flows gives the complete base flow for $\beta \ll 1$. For high inertia ($\beta \gg 1$), the fluid flow throughout the channel is predominantly inviscid. This inviscid flow is driven only by the normal component of the motion of the rotating walls, which periodically 'pushes and pulls' on the fluid, producing an extensional, irrotational flow (Fig. 1b-iii).

When a particle is placed in the base flow, a disturbance flow is generated (Fig. 1c) that produces a measurable signal. The two flows altogether satisfy the no-penetration and no-slip conditions at the particle's surface. The disturbance flow alters the torque exerted by the fluid on the channel walls. This change in torque modifies the effective rigidity of the cantilever, shifting its resonant frequency by $\Delta f_n$. We developed a theory for $\Delta f_n$, for a particle of scaled radius, $R = a/H$ where $a$ is its dimensional radius, and dimensionless mass density, $\gamma = \rho_p/\rho$ where $\rho_p$ is the density of the particle (supplementary note 1):

$$\Delta f_n = -f\, \alpha_v(\beta\,|\,\gamma, R, z)\, \mathcal{V} \left(\frac{dW}{dx}\right)^2, \quad (1)$$

where $f$ is the resonant frequency when no particle is present, $\mathcal{V} = \rho V^{5/3}/\left(2\,[6\pi^2]^{1/3} m_{\text{eff}} L^2\right)$ is a dimensionless particle volume factor with particle volume $V$, $z$ is the vertical position of the particle within the channel scaled by the channel height, $H$, $L$ is the cantilever length, $m_{\text{eff}}$ is the effective mass of the cantilever and $W(x)$ is the displacement mode shape of the cantilever along its length $x$, which is scaled by $L$. The viscous enhancement factor, $\alpha_v$, defines the relative contribution of fluid viscosity to the frequency shift; $\alpha_v \to 1$ for inviscid flow ($\beta \to \infty$, Fig. 2). The disturbance flow in this inviscid limit is identical to the flow derived from the scattered acoustic field at large wavelength, previously studied computationally.[19]

While $\Delta f_n$ depends on both $\mathcal{V}$ and $\alpha_v$, information on the rich variety of flow behavior is contained within $\alpha_v$; $\mathcal{V}$ depends only on the particle's volume and properties of the cantilever. Specifically, $\alpha_v$ exhibits a strong non-monotonic dependence on $\beta$ (Fig. 2), and a conditional dependence—with respect to $\beta$—on particle density, $\gamma$ (Fig. 2), radius, $R$ (Fig. S3) and $z$-position (Fig. S4). Because the general case for arbitrary $\beta$ is complicated, we gain understanding by again exploring the limits of low ($\beta \ll 1$) and high ($\beta \gg 1$) inertia, for particles placed at the channel center ($z = 0$).

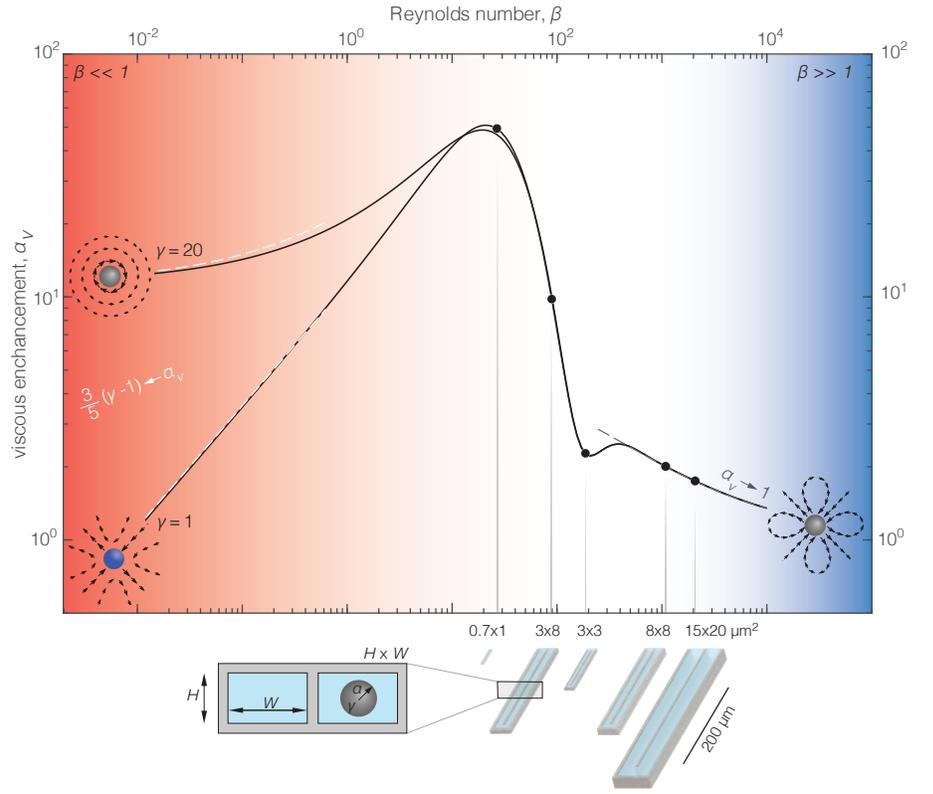

**Figure 2 | Viscous enhancement of frequency change due to local rotation.** The theoretically derived $\alpha_v$ expressing the viscous enhancement of signal $\Delta f_n$, is shown for $R \equiv a/H = 0.1$ where $a$ is the dimensional radius of the particle. Similar non-monotonic behavior vs $\beta$ is observed for other $a/H$ ratios (Fig. S3). Ratio of the particle density to the fluid density is $\gamma = \rho_p/\rho$. Dashed lines represent the asymptotic limits for $\beta \ll 1$ and $\beta \gg 1$. Black circles represent the cantilevers (bottom right) used in the experiments; $H \times W$ gives the dimensions of the cross-sectional area of the flow channel (bottom left). Schematics of the flow fields correspond to the disturbance flows in Fig. 1c.

For low inertia ($\beta \ll 1$), the particle experiences the primary base flow of rigid-body rotation (Fig. 1b, i). Although fluid inertia is negligible, the particle's inertia causes it to rotate relative to the base flow, when particle density differs from that of the fluid ($\gamma \neq 1$). This generates a rotational disturbance flow (Fig. 1c, i, Movie 1). Notably, in the limit $\beta \to 0$, the viscous enhancement factor, $\alpha_v$, is constant and determined only by $\gamma$ (Fig. 2, $\beta \ll 1$). While a disturbance flow due to the (secondary) shear flow also exists, it is significant only for near density-matched particles ($\gamma \approx 1$) and the sole contribution for $\gamma = 1$. Such particles do not rotate relative to the base flow, and thus cannot react to the primary base flow (Fig. 1b, i) or the rotational component of this secondary flow (Fig. S1a, iv). In such cases, a disturbance flow is generated by the secondary flow (Fig. 1a, v). This flow 'pushes and pulls' on the particle surface, generating a viscous quadrupole disturbance flow (Fig. 1c, ii, Movie 2). Because the secondary flow vanishes in the limit $\beta \to 0$, so does the viscous enhancement factor ($\alpha_v \to 0$ for $\gamma = 1$, Fig. 2, $\beta \ll 1$).

For any particle that is not precisely or near density-matched ($\gamma \not\approx 1$), the viscosity-dominated rotating base flow activates the rotational inertia of the particle, converting it into a 'viscous flywheel'. While negligible inertia exists in the fluid, and there is no vertical displacement of the particle, a frequency change, $\Delta f_n$, occurs that depends on the particle mass density.

For high inertia ($\beta \gg 1$), the particle experiences an extensional base flow (Fig. 1b-iii). In contrast to $\beta \ll 1$, this base flow is inviscid, and thus irrotational, despite being driven by the rotating walls. Here, the particle does not rotate, regardless of its density, $\gamma$, and generates a quadrupole disturbance flow (Figs. 1c-iii, 2, $\beta \gg 1$, Movie 3) similar to the density-matched particle above (Fig.1c-ii). As a result of this irrotational base flow, the viscous enhancement factor, $\alpha_v$, is independent of particle density, $\gamma$.

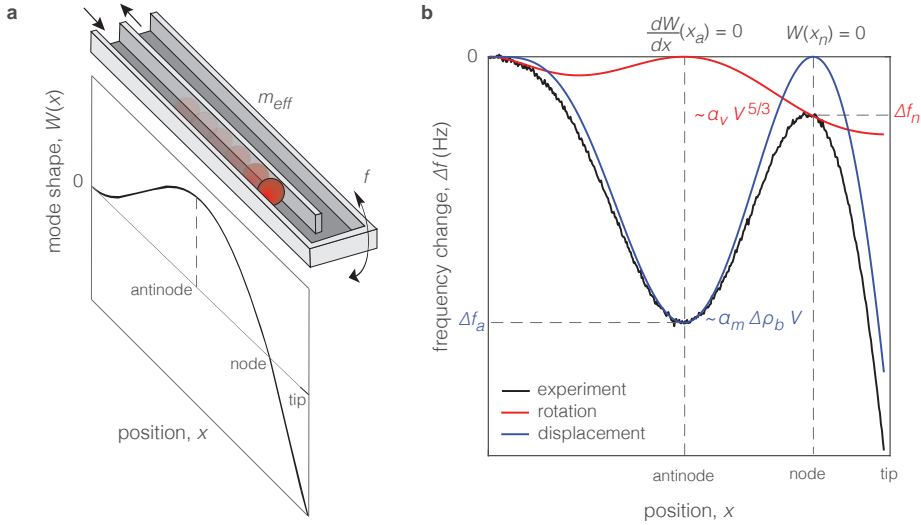

**Figure 3 | Experimental realization of local rotation in microcantilevers. a**, Schematic of a cantilever, with effective mass, $m_{eff}$, driven in its second flexural mode, with resonant frequency $f$; $W(x)$ specifies the displacement along the length of the cantilever, $x$. At the node (i.e., where $W(x_n) = 0$), there is local rotation only, $dW/dx|_{x_n} \neq 0$. **b**, Frequency change, $\Delta f$, induced by a particle with nonzero buoyant mass, $\Delta m$, and volume $V$. The experimental signal, $\Delta f$ (black), consists of two signals: one due to displacement (blue) and one due to rotation (red). The parameter $\alpha_m$ represents the mass discrepancy in the displacement signal.[22]

In the high $\beta$ limit, the thin viscous boundary layers around the particle and in the vicinity of the walls do not overlap. Decreasing $\beta$, i.e., increasing the viscous penetration depth, $\delta$, results in a monotonic increase to $\alpha_v$ (Fig. 2, $\beta \gg 1$); due to an increase in the effective particle size. Ultimately, this decrease in $\beta$ causes these viscous boundary layers to overlap, producing a non-monotonic variation of $\alpha_v$ (Fig. 2, $\beta \approx 200$). This overlap, with its resulting effect on $\alpha_v$, is initiated at different values of $\beta$, which depend on both the particle's radius, $R$, (Fig. S3) and its $z$-position (Fig. S4).

To validate the theory, we experimentally measured the resonant frequency change, $\Delta f$, of six types of SMRs/SNRs cantilevers (Table S1) while flowing polystyrene and glass nanoparticles through their channels containing water (Table S2). The dimensions of the cantilevers differ significantly, with effective masses, $m_{eff}$, spanning three orders-of-magnitude (Fig. S5), while their Reynolds numbers range from $\beta \approx 30$ for the smallest device, to $\beta \approx 2,000$ for the largest (Fig. 2, black points). The smaller cantilevers exhibit increased mass responsivity while being more susceptible to the effects of fluid viscosity.[20] We actuated the cantilevers at their second flexural mode (Fig. 3a) because this mode provides the lowest frequency at which a vibrational node occurs.[17] When a particle flows through the cantilever channel, a frequency shift signal is measured containing contributions from (i) the SMR/SNR's vertical displacement, which is standard in inertial sensing,[14] and (ii) its rotational dynamics (Fig. 3b). To isolate the contribution from rotation, i.e., $\Delta f_n$ in Eq. (1), we processed the total measured signal, $\Delta f$, using an iterative algorithm (Fig. S6).

Theory predicts an inconsequential dependence of $\alpha_v$ on particle density, $\gamma$, for $\beta \gtrsim 10$ (Figs. 2, S3, S4), indicating that $\Delta f_n$ in Eq. (1) depends experimentally on particle volume only. To test this prediction, we conducted two series of experiments.

First, we flowed polystyrene calibration particles with nominal radii in the range $a = 125 - 6,000$ nm, and measured $\Delta f_n$ for each particle (Fig. 4a, Movie 4). Next, we determined the volume, $V_{\text{meas}}$, of each particle by fitting the experimental measurements to Eq. (1) using an iterative algorithm (Fig. S6). We also independently measured each particle's volume, $V_{\text{ref}}$—termed the "reference volume"—from the buoyant mass, extracted using the antinode signal[21] $\Delta f_a$, (Fig. 3b). This used the known density of the particles ($\rho_{p,\text{pol}} = 1,050$ kg/m$^3$) and that of the surrounding fluid ($\rho = 997$ kg/m$^3$), respectively. We observed excellent agreement between $V_{\text{meas}}$ and $V_{\text{ref}}$ over the entire experimental dataset (Fig. 4b).

Experimental measurement of particle volume has three main sources of error (Supplementary Note 2), including: (i) non-linear error propagation when determining particle volume from the signal, (ii) effect of the channel walls on the disturbance flow not being accounted for by the theory, and (iii) error associated with assuming that the particles are at the channel center ($z = 0$) when analyzing the experimental data. Using Monte-Carlo simulations (Figs. S7, S8), signal-to-noise calculations (Figs. S9, S10) and scaling arguments (Supplementary Note 2), we found that these errors are minimized for $\beta \gtrsim 100$ and $R > 0.25$. Further improvement may be obtained by actuating the cantilevers at multiple modes simultaneously,[21] which would enable determination of the particle $z$-position.

Second, we flowed glass particles of similar volume to the polystyrene particles but with a buoyant density, $\Delta \rho_p = \rho_p - \rho$, that is one order-of-magnitude larger. Analysis[22] of the measured antinode signal, $\Delta f_a$, reveals an enhanced buoyant density for these glass particles of commensurately increased magnitude (Fig. S11a,b). Even so, analysis of the measured rotation signal, $\Delta f_n$, directly gives a particle volume, $V_{\text{meas}}$, of similar magnitude to the polystyrene particle volumes (Fig. S11c,d). This shows that the rotational signal, $\Delta f_n$, is independent of particle mass and gives direct access to particle volume.

This finding enables a unique, direct measurement of particle density. Combining measurement of the particle's volume, with that of its buoyant mass, respectively using $\Delta f_n$ and $\Delta f_a$ from a single pass of a particle through the cantilever, we measured particle density with an accuracy of at least 99% for particles that are larger than half the height of the microchannel (Fig. S12). Previous methodologies based on fluid-filled cantilevers rely on complex fluid exchanges to measure particle mass and volume; this limits throughput to <6 particles per minute.[23,24] In comparison, the present methodology extracts these properties simultaneously from a simple, single measurement at a throughput that is 10-fold greater.

Overall, the realization of rotational inertial sensing defines a new paradigm in inertial sensing for characterizing the vibrational response of fluid-suspended micro- and nanoparticles. Interaction of the particle within a rotating fluid-filled microchannel leads to a rich array of flow mechanisms enabling this new sensing modality. The presented analytical theory directly augments existing theory for inertial sensing that use fluid-filled cantilever and plates.[6] This rigorously accounts for the ubiquitous—yet previously complicating and ignored—effects of rotation.

## Methods

**Fabrication and design of devices.** The nanochannel and microchannel suspended resonator devices (SNR, SMR) were fabrication at Innovative Micro Technology (Santa Barbara, CA, USA) and CEA-LETI (France) using 6-inch and 8-inch silicon wafer technology[6,16,25]. The technology enables the cantilevers of each device to oscillate in a dedicated vacuum cavity containing an on-chip getter to maintain the high vacuum, thus ensuring high quality factor during operation. Each device (Table S1) has either one cantilever (devices 0.7x1.0, 3x3) or two cantilevers (devices 3x8, 8x8, 15x20a, 15x20b). For each cantilever, there are four fluidic ports drilled on the top glass wafer to access two bypass channels respectively connected to the inlet and the outlet of each cantilever (Fig. 3a).

**Operation of devices.** Each SMR/SNR device was actuated at the second vibration mode (Fig. 3a) by a piezo-ceramic plate on top of which the device was epoxy-bonded, using a dedicated phase-locked loop (PLL) in closed loop[21]. Precision pressure regulators (electronically controlled Proportion Air QPV1 and manually controlled Omega PRG101-25) were used to flow particle solutions within each device. To measure the signal of change in resonance frequency, $\Delta f$, (Fig. 3b), either optical[26] (0.7x1, 3x3, 3x8, 8x8) or piezoresistive readout[27,28] (15x20a,b) methods were employed, while in both cases a field programmable gate array (FPGA, Altera Cyclone IV on DE2-115) was used, connected via ethernet cable to a desktop computer. To ensure adequate sampling[29] of $\Delta f$, the transit time $\Delta t_{transit}$ of the particle through the cantilever was set such that $\Delta t_{transit}(sec) > 24/bw(Hz)$, where $bw$ is the bandwidth of the PLL loop. The experiments were performed using a custom code written in LabVIEW 2017 software.

**Preparation of particle solutions.** The polystyrene calibration particles (Table S2) were originally supplied by the vendor in aqueous solutions with concentrations of 0.20-1.00 % solids. For flowing into the cantilevers, they were diluted by a factor of 500-1000 times using purified, filtered (filter size 20 nm) water. Occasionally, to prevent pinning of polystyrene particles inside the cantilever, Tween 20 (Sigma Aldrich, MO USA) was added at a percentage of approximately 0.05% per volume of particle solution. The glass particles were originally supplied by vendor in dry form. For flowing into the cantilevers, they were suspended in purified, filtered water at concentrations of approximately 0.03 mg/μL with 0.0003% Tween 20. In the event of clogging, the cantilever device was flushed with a sequence of water, isopropanol, acetone, toluene until device was unclogged.

**Post processing of experimental data.** The experimental data were analyzed by implementing an iterative algorithm (Fig. S6) in MATLAB 2019b.

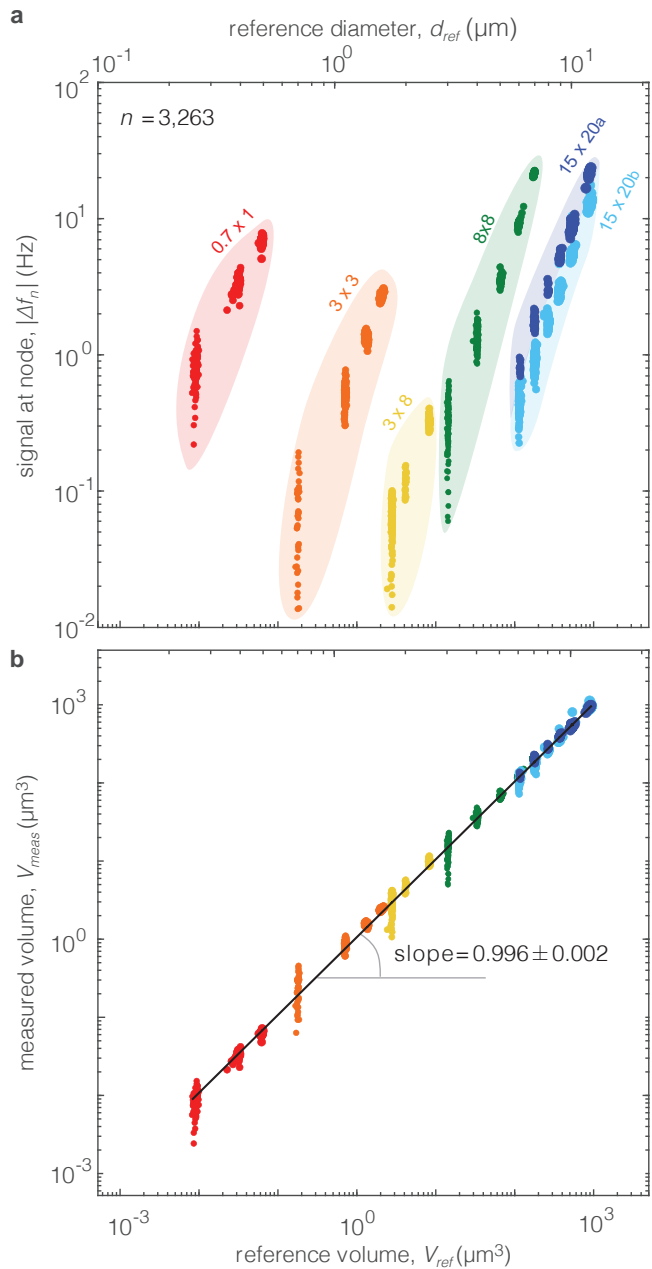

**Figure 4 | Experimental validation of theory. a**, Measurements of the frequency change signal $\Delta f_n$ at the node position (Fig. 3b, red) for polystyrene particles suspended in water in six different devices. The reference size, (radius, $a_{ref}$, and volume, $V_{ref}$) are measured from the antinode signal. Note that in the present experiments, $\Delta f_n < 0$. The colored islands represent each different device with symbols referring to Fig. 2; $n$ is the total number of measurements across all experiments. **b**, Measured volume, $V_{meas}$, vs reference particle size as in **a**. $V_{meas}$ is calculated from the node signal, $\Delta f_n$, using Eq. (1) (Fig. S6). A linear regression of $V_{meas}$ vs $V_{ref}$ is performed where the bounds give a 99.9% confidence interval.

**Acknowledgements**

We thank Iris E. Hwang and John H. Kang for helpful comments. S.R.M. and G.K. acknowledge support from the Ludwig Center for Molecular Oncology. J.F.C. and J.E.S. acknowledge support from the Australian Research Council Centre of Excellence in Exciton Science (CE170100026) and the Australian Research Council Grants Scheme.

**Author contributions**

G.K., J.F.C., J.E.S. and S.R.M conceived the study. G.K. and S.R.M. designed the experiments. G.K. carried out the experiments with the exception of the experiments with the 3x3 device, which S.M.K. performed. V.A. designed and provided the 0.7x1.0 device. J.F.C. and J.E.S. developed the theory which built on an alternate inviscid theory developed by G.K. G.K. and J.F.C. analyzed the data and performed the Monte-Carlo simulations. G.K., J.F.C., J.E.S. and S.R.M. wrote the paper with V.A. providing input.

**Competing interests**

S.R.M. is a co-founder of Travera and Affinity Biosensors, which develops technologies relevant to the research presented in this work.

# Supplementary information

# Viscous and flywheel sensing of nanoparticles

Georgios Katsikis,[a,b] Jesse F. Collis,[a,c] Scott M. Knudsen,[b] Vincent Agache,[b,d] John E. Sader,[c,e] Scott R. Manalis[b,f,g,h]

[a] These authors contributed equally to the work.
[b] Koch Institute for Integrative Cancer Research, Massachusetts Institute of Technology, Cambridge, MA 02139 USA.
[c] ARC Centre of Excellence in Exciton Science, School of Mathematics and Statistics, The University of Melbourne, VIC 3010, Australia.
[d] Université Grenoble Alpes, CEA, LETI, 38000, Grenoble France.
[e] Corresponding author: jsader@unimelb.edu.au.
[f] Department of Biological Engineering, Massachusetts Institute of Technology, Cambridge, MA, 02139 USA.
[g] Department of Mechanical Engineering, Massachusetts Institute of Technology, Cambridge, MA, 02139 USA.
[h] Corresponding author: srm@mit.edu.

# Supplementary Figures

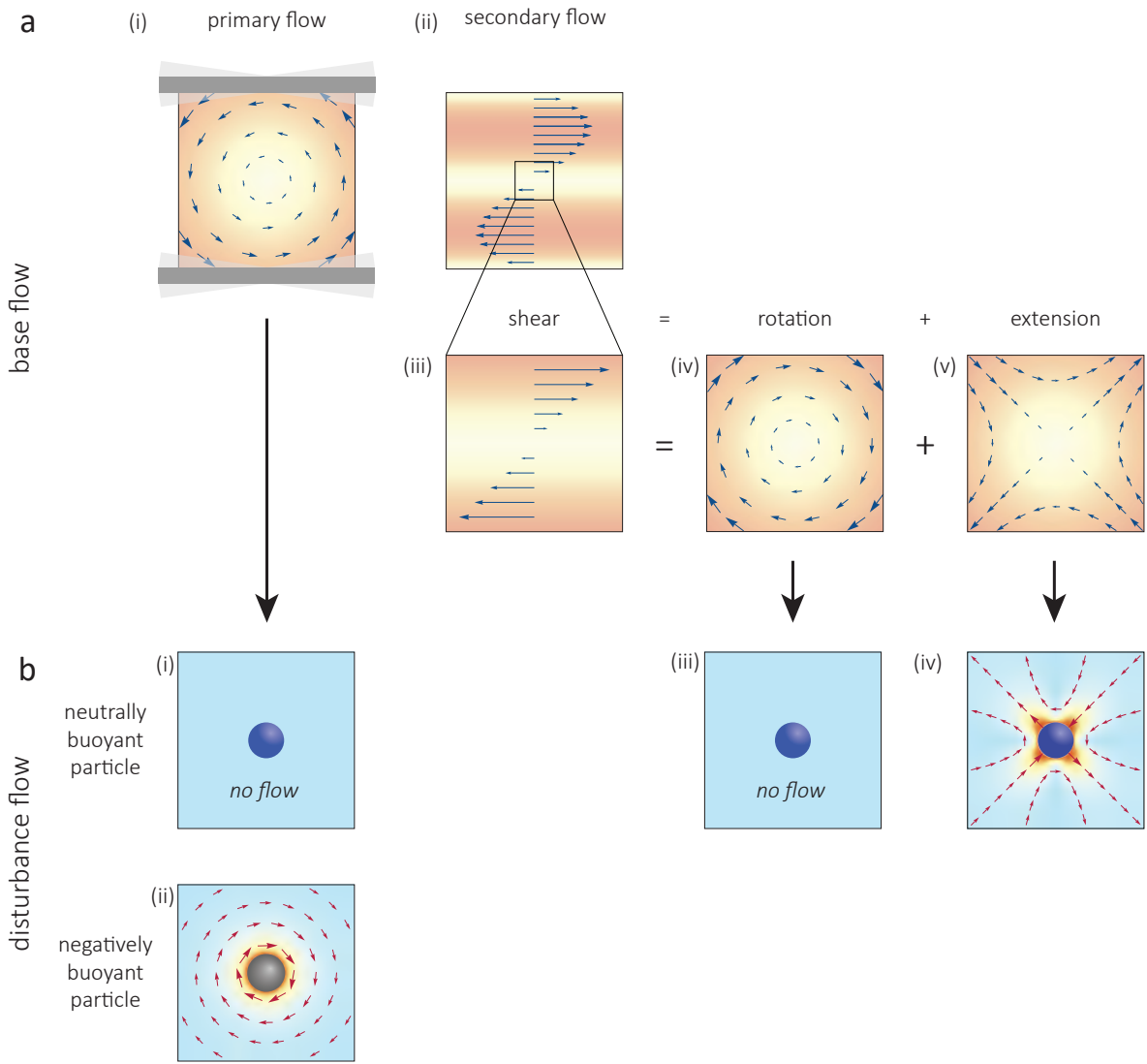

**Figure S1 | Illustration of the flow within a rotating microchannel at low Reynolds number. a,** Flow fields generated within the rotating microchannel in the absence of a suspended particle. (i) The flow is a rigid body rotation in the limit of zero inertia. (ii) Infinitesimal inertia causes the fluid to lag its primary flow counterpart. This produces a secondary, non-linear, shear flow. (iii) As the theory assumes the suspended particle is much smaller than the channel height ($a \ll H$), this non-linear shear flow is approximated as linear over the length scale of the particle. (iv, v) A pure (linear) shear flow is decomposable into a pure rotation and a pure extension, i.e., $z\hat{x} = (z\hat{x} - x\hat{z})/2 + (z\hat{x} + x\hat{z})/2$. **b,** Flow due to a particle suspended in the base flows in (a). (i) A neutrally buoyant particle rotates with the same angular velocity as the base flow and therefore produces no disturbance flow. (ii) A negatively buoyant particle produces a rotational disturbance flow and is discussed further in the caption of Fig. 1. (iii) The neutrally buoyant particle does not react to the rotational component of the secondary shear flow for the same reason as (i). (iv) It does, however, react to the shear flow's extensional component, producing a quadrupole disturbance flow (iv), similar to the case for high inertia in Fig. 1. The difference between (iv) here and Fig. 1c-ii is that the flow around the particle is dominated by viscosity in the former and inertia in the latter.

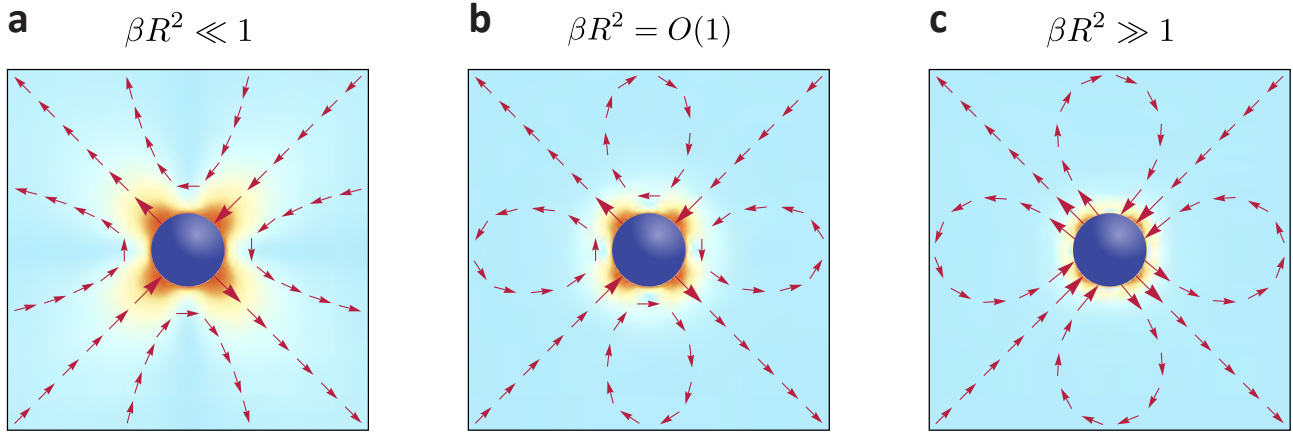

**Figure S2 | Quadrupole disturbance flows (i.e., particle reaction to an extensional base flow) for increasing particle/viscous penetration depth ratio. a**, For $\beta R^2 \ll 1$, the disturbance flow generated by the particle is viscous everywhere and decays at a rate of $1/r^2$ where $r$ is the non-dimensional distance from the particle surface. The streamlines are always open, i.e., there are no recirculation zones. This regime results in the strongest disturbance flow of the three results presented here, and hence also the largest magnitude of $\alpha_v$. **b**, For $\beta R^2 = O(1)$, the disturbance flow is viscous dominated from the particle surface to a distance of $O(1)$ from the particle surface; the flow decays at a rate of $1/r^2$ in this region. Outside this region, the flow is primarily inviscid, where it decays at a rate of $1/r^4$. This results in a weaker disturbance flow than (a), and hence a reduction to the magnitude of $\alpha_v$. **c**, For $\beta R^2 \gg 1$, the disturbance flow is primarily inviscid and so decays at a rate of $1/r^4$ from the particle surface. This results in a weaker disturbance flow than (b) and hence the smallest magnitude of $\alpha_v$ for the three results presented here.

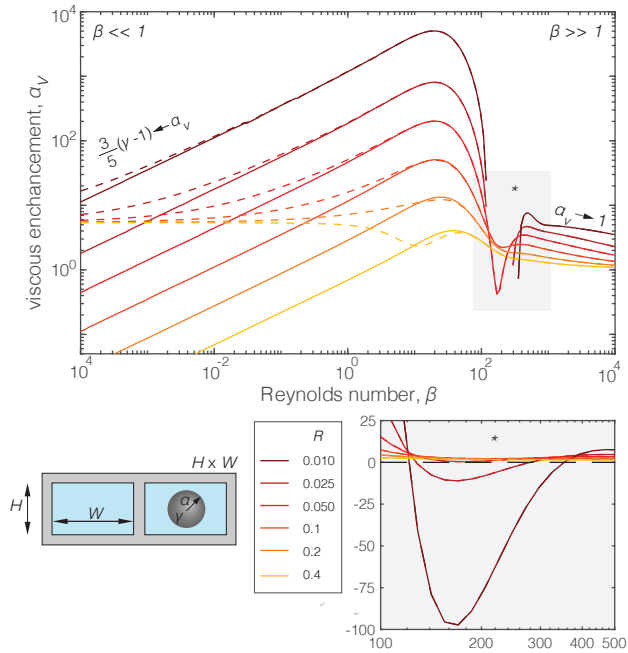

**Figure S3 | Viscous enhancement factor, $\alpha_v$, vs Reynolds number $\beta$ for varying ratios of particle radius to channel height ($R = a/H$).** Assuming that the particle is located at the center of the channel ($z = 0$), for most Reynolds numbers, decreasing particle size, and hence increasing the effects of viscosity in the vicinity of the particle (Fig. S2), increases the magnitude of $\alpha_v$. At intermediate $\beta$, the complex interplay between the particle size, and the viscous boundary layers in the vicinity of the particle and the channel walls, results in a sensitive dependence of $\alpha_v$ on $R$, even reversing the sign of $\alpha_v$ (gray inset with asterisk). The ratio of particle density over fluid density, $\gamma = \rho_p/\rho$, is shown with continuous lines for $\gamma = 1$ and dashed lines for $\gamma = 10$. The curve for $R = 0.1$ is identical to the corresponding curve in Fig. 2.

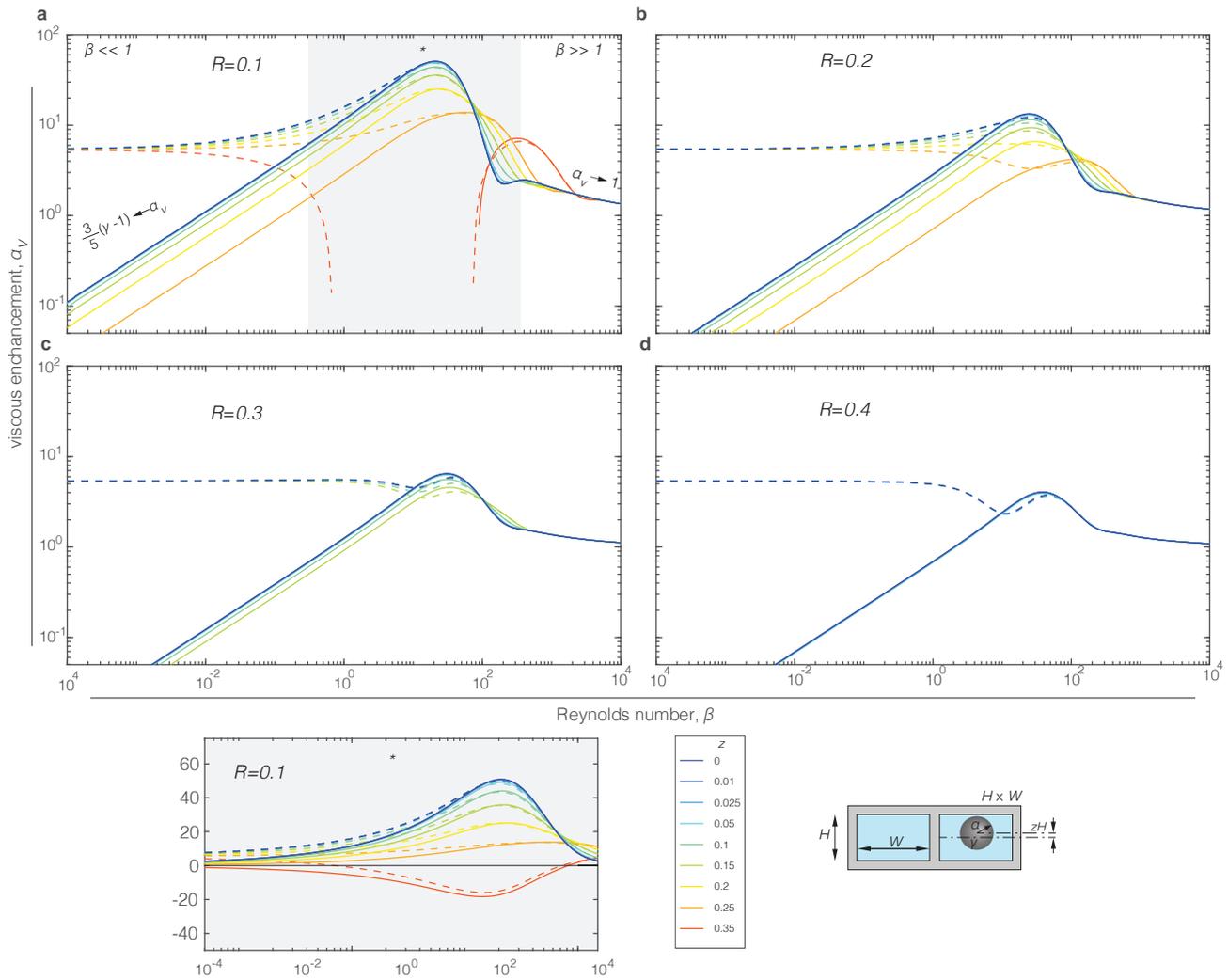

**Figure S4 | Viscous enhancement factor, $\alpha_v$, vs Reynolds number, $\beta$, for varying z-position. a-d,** Four different radii, $R = 0.1, 0.2, 0.3, 0.4$, respectively shown in **a-d**, for z-positions in the range $0 - 0.35$. Radii are defined as $R = a/H$ where $a$ is the dimensional radius of the particle. Only positive $z$ is shown here as $\alpha_v$ is symmetric in $z$. The limits $\beta \ll 1$ and $\beta \gg 1$, and the corresponding asymptotic formulae, are shown only in (a) but apply to all panels. As a particle cannot cross over a wall, the maximum z-position is given by $z_{max} = 0.5 - R$, hence the decreasing number of curves from (b)-(d). The ratio of particle density over fluid density $\gamma = \rho_p/\rho$ is shown with continuous lines for $\gamma = 1$ and dashed lines for $\gamma = 10$. For $R = 0.1$, at intermediate $\beta$, $\alpha_v$ reverses sign and so we display a log-linear plot for the gray inset (asterisk). The curve for $R = 0.1$ and $z = 0$ is identical to the corresponding curve in Fig. 2. Note that, while the theory is derived in the limit $R \ll 1$, we find good agreement between theory and experiment up to the experimental maximum of $R = 0.4375$.

| Schematic of SNR/SMR cantilever devices |
|---|

| Dimensions (μm) | | | | | | |
|---|---|---|---|---|---|---|
| name | 0.7x1 | 3x3 | 3x8 | 8x8 | 15x20a | 15x20b |
| $H$ | 0.7 | 3 | 3 | 8 | 15 | 15 |
| $W$ | 1.0 | 3 | 8 | 8 | 20 | 20 |
| $L$ | 32.5 | 100 | 210 | 210 | 321 | 355 |
| $H_{bot}$ | 0.2 | 1 | 2 | 2 | 2 | 2 |
| $H_{top}$ | 0.2 | 1 | 2 | 2 | 2 | 2 |
| $W_{ext}$ | 1.0 | 2 | 6 | 6 | 6 | 6 |
| $W_{int}$ | 0.5 | 1.5 | 5 | 5 | 5 | 5 |
| $L_{wall}$ | 31.0 | 98 | 196 | 196 | 295 | 329 |
| $L_{tip}$ | 0.5 | 1 | 6 | 6 | 6 | 6 |
| Baseline frequency, $f$ (MHz), Reynolds number, $\beta$, Quality factor, $Q$ | | | | | | |
| $f$ | 7.54 | 3.06 | 1.46 | 2.37 | 1.34 | 1.19 |
| $\beta$ | 26 | 194 | 92 | 1,067 | 2,120 | 1,882 |
| $Q$ | 3,000 | 1,800 | 4,000 | 1,600 | 800 | 1,000 |

**Table S1 | Characteristics of the cantilevers used in this study.** Schematic of a cantilever with its geometric dimensions. The cantilever in the perspective view is shown without its top lid for clarity; the cantilever has a top lid with thickness $H_{\text{top}}$ as shown in the cross-section view of the flow channel. All cantilevers have the same geometrical features. The table gives the dimensions of each cantilever and their corresponding baseline frequency, $f$, Reynolds number, $\beta$, and quality factor, $Q$, when operated in 2$^{\text{nd}}$ resonant mode while filled with water. The bolded dimensions ($H$, $L$, $f$ and $\beta$) are used directly in Eq. (1) to calculate the change in resonant frequency, $\Delta f$.

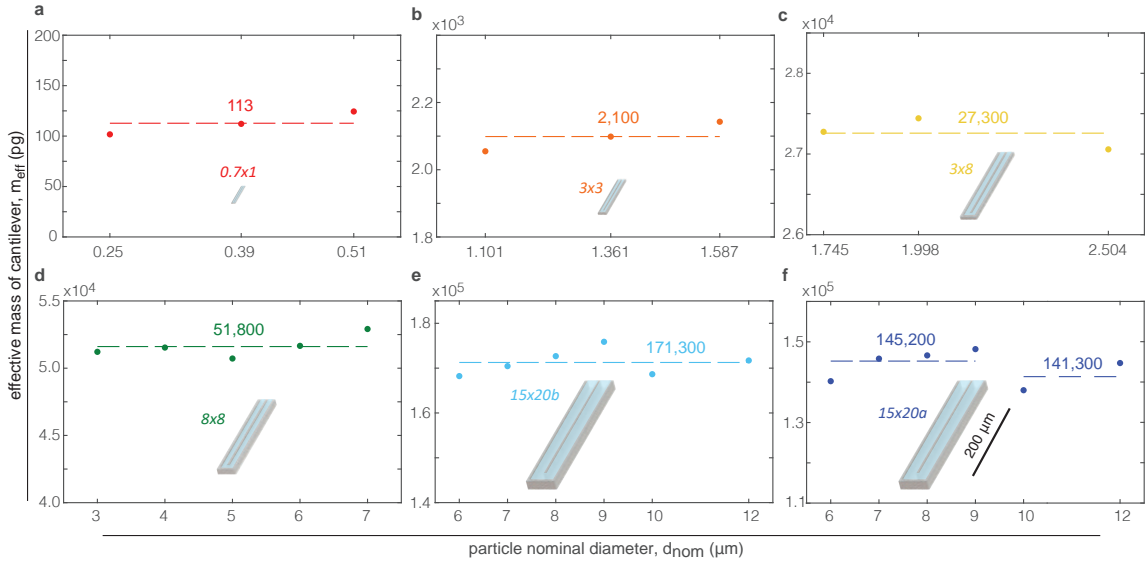

**Figure S5 | Calculation of the effective mass, $m_{eff}$ for each cantilever used in this study. a-f,** Code names and colors are the same as in Figs. 2, 4. For each cantilever, $m_{eff}$ is calculated by applying Eq. (20) (supplementary note 1) at the antinode position (Fig. 3b) where $\Delta f_a$ and $f_0$ are measured experimentally. The buoyant mass, $m_b$, is calculated as $m_b = (\rho_p - \rho)\pi d_{nom}^3/6$ where $d_{nom}$ is the nominal diameter of a given monodisperse solution of polystyrene calibration particles and $\rho = 1,050$ kg/m$^3$ is their density (Table S2). The dashed lines represent the mean value of $m_{eff}$ calculated from solutions of different $d_{nom}$ in each cantilever. Two cantilevers of type 15x20a were used.

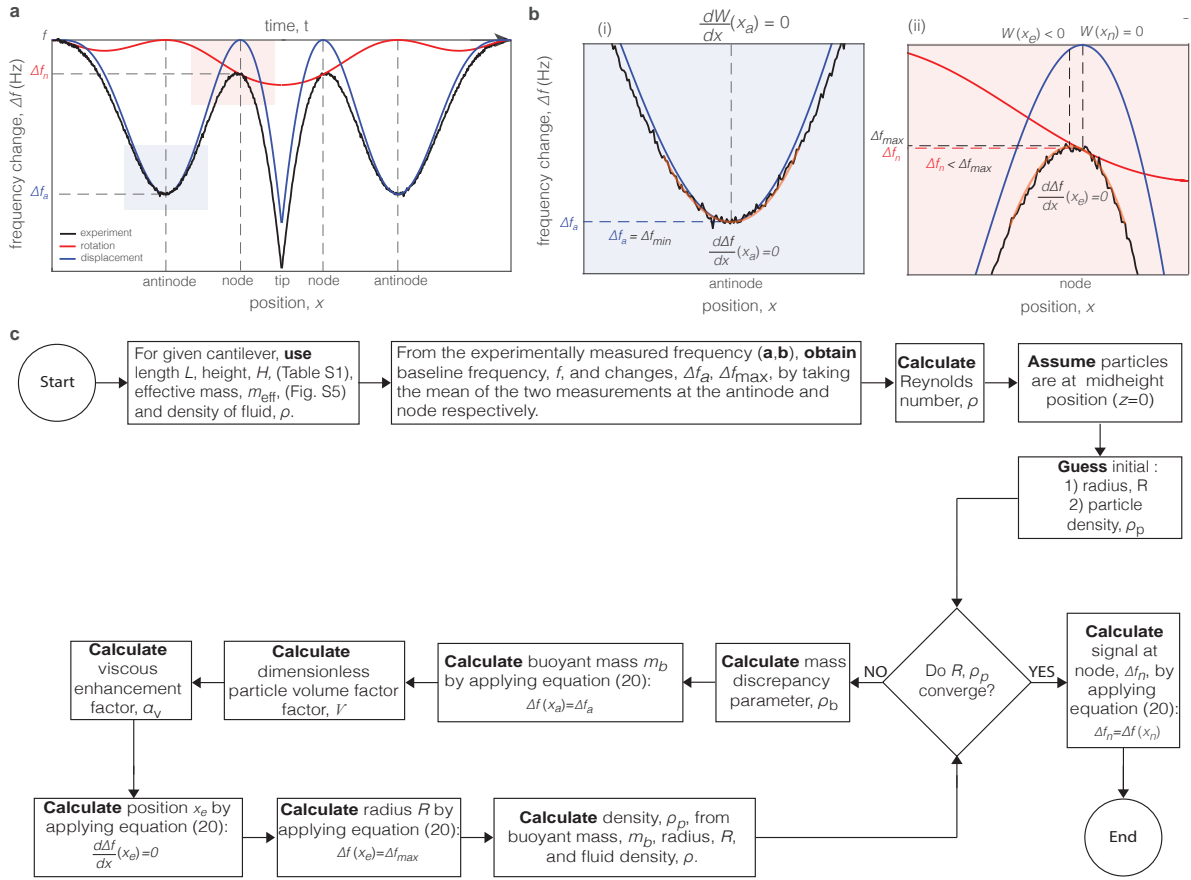

**Figure S6 | Extraction of signals from measured resonant frequency changes and corresponding iteration algorithm for calculating particle properties. a,** Frequency change, $\Delta f$, induced by a particle vs $x$-position along the length of the cantilever. The experimental signal, $\Delta f$ (black), consists of two signals: the first due to displacement (blue) and the second due to rotation (red), similar to Fig. 3b but shown here for the double measurement, i.e., the particle transit down each side of the cantilever (Fig. 3a). **b,** Frequency change, $\Delta f$, induced in the vicinity of the (i) antinode and (ii) node with background colors corresponding to areas in (a). The orange curves represent the smoothed curves for extracting the local (i) minimum and (ii) maximum of $\Delta f$. Note that the $x$-position of the local maximum in the black curve of (ii) does not exactly correspond to the $x$-position of the node. **c,** Flow chart of the iterative algorithm used to calculate the particle properties.

| nominal diameter $d_{nom}$ (µm) | 0.25 | 0.39 | 0.51 | 0.7 | 1.1 | 1.3 | 1.6 | 1.8 | 2 | 2.5 |
|---|---|---|---|---|---|---|---|---|---|---|
| diameter (mean+error in µm) | not available | | | 0.7086+0.0235 | 1.101+0.017 | 1.361+0.015 | 1.587+0.025 | 1.745+0.025 | 1.998+0.022 | 2.504+0.025 |
| size distribution (µm) | | | | - | 0.012 | 0.021 | 0.021 | 0.019 | 0.020 | 0.025 |
| coefficient of variation, CV (%) | 5 to 10 | | | 3 | 1.1 | 1.5 | 1.3 | 1.1 | 1.0 | 1.0 |
| density (g/cm$^3$) | 1.05 (polystyrene) | | | | | | | | | |
| catalog number | PS02N | PS02N | PS03N | NT12N | 4011A | 4013A | 4016A | 4018A | 4202A | 4025A |
| lot number | 7307 | 6703 | 5970 | 7141 | 44653 | 44330 | 35209 | 35306 | 34646 | 35261 |
| vendor | Bangs Laboratories | | | Thermo Scientific Duke Standards | | | | | | |

| nominal diameter $d_{nom}$ (µm) | 3 | 4 | 5 | 6 | 7 | 8 | 9 | 10 | 12 |
|---|---|---|---|---|---|---|---|---|---|
| diameter (mean+error in µm) | 3.005+0.027 | 4.000+0.043 | 5.000+0.042 | 6.007+0.040 | 6.976+0.057 | 7.979+0.075 | 8.956+0.082 | 10.00+0.05 | 12.01+0.11 |
| size distribution (µm) | 0.029 | 0.04 | 0.05 | 0.06 | 0.07 | 0.09 | 0.09 | 0.09 | 0.12 |
| coefficient of variation, CV (%) | 1.1 | 1.0 | 1.0 | 1.0 | 1.0 | 1.1 | 1.0 | 0.9 | 1.0 |
| density (g/cm$^3$) | 1.05 (polystyrene) | | | | | | | | |
| catalog number | 4203A | 4204A | 4205A | 4206A | 4207A | 4208A | 4209A | 4210A | 4212A |
| lot number | 34742 | 203429 | 214115 | 204557 | 188878 | 36698 | 43716 | 36342 | 44085 |
| vendor | Thermo Scientific Duke Standards | | | | | | | | |

| nominal diameter $d_{nom}$ (µm) | 9.2 |
|---|---|
| diameter (mean+error in µm) | not available |
| size distribution (µm) | |
| coefficient of variation, CV (%) | 4.8 |
| density (g/cm$^3$) | 2.0 (silica) |
| catalog number | SiO2MS-2.0 9.2um |
| lot number | 171115-1048 |
| vendor | Cospheric |

**Table S2 | Manufacturer specifications for all particles used in this study**.

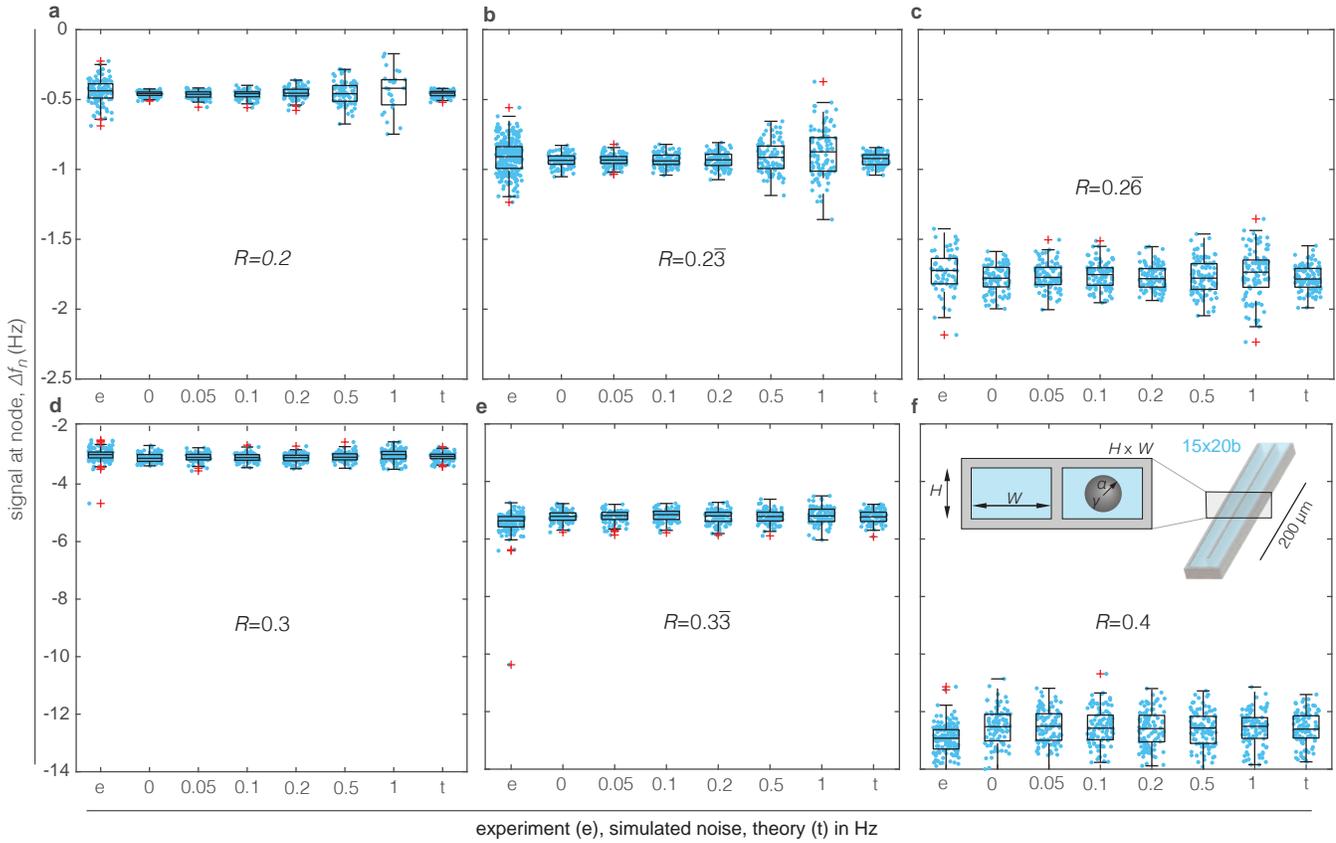

**Figure S7 | Comparison of the node signal from experiments with Monte-Carlo simulations at varying noise levels**. **a-f**, Node signal, $\Delta f_n$, for particles of nominal diameter $d_{nom} = 6 - 12$ µm (corresponding to $R = 0.2 - 0.4$) in the 15x20b device (Fig. 4, 15x20b). The inset in (f) shows the cross section of the flow channel with area $H \times W$. Noise simulations consist of theoretically calculated $\Delta f_n$ using Eq. (1) with added Gaussian noise of standard deviation $\sigma = 0, 0.05, 0.1, 0.2, 1$ Hz with $n = 100$ individually simulated points. Frequency changes, $\Delta f$, from experiments (symbol e) and noise simulations (horizontal axis numbers have the units Hz) are processed with the same algorithm for extracting $\Delta f_n$ from experiments (Fig. S6). Signals of $\Delta f_n$ from either experiments or simulations are rejected if the signal-to-noise ratio is low (for example in (a), 1 Hz). The (noiseless) theory (symbol t) is calculated directly from Eq. (1). The general trend observed here is that noise dominates the signal for small particles ($d_{nom} \leq 7$ µm or $R \leq 0.2\bar{6}$), and polydispersity dominates the signal for larger particles. Boxplots have similar notation as in Fig. S7.

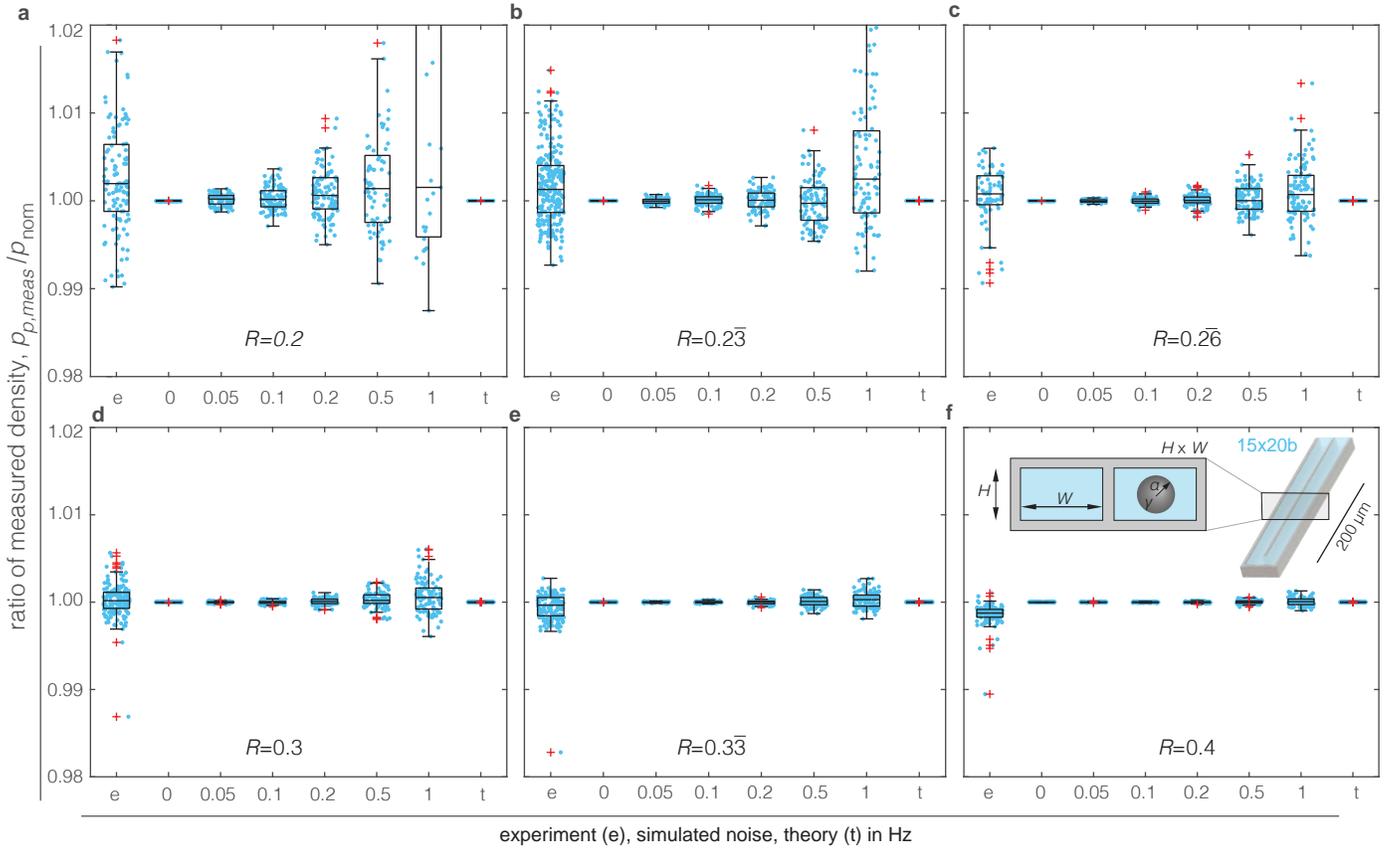

**Figure S8 | Comparison of microparticle density calculated from experiments, with Monte-Carlo simulations varying the noise level**. **a-f**, Measured density ratios $\rho_{p,\text{meas}}/\rho_{\text{nom}}$ of particles of nominal diameter $d_{\text{nom}} = 6 - 12$ μm (corresponding to $R = 0.2 - 0.4$) in the 15x20b device (Fig. 4, 15x20b). The data, insets, numbers and box plot symbols are the same as in Fig. S7. As $R$ increases, $\rho_{p,\text{meas}}/\rho_{\text{nom}} \rightarrow 1$ due to a higher signal to noise ratio. However, for the maximum diameter $d_{\text{nom}} = 12$ μm ($R = 0.4$), there is a deviation of the experimental data (symbol e) from $\rho_{p,\text{meas}}/\rho_{\text{nom}} = 1$ which is likely due to the effects of finite particle size. Boxplots have similar notation as in Fig. S7.

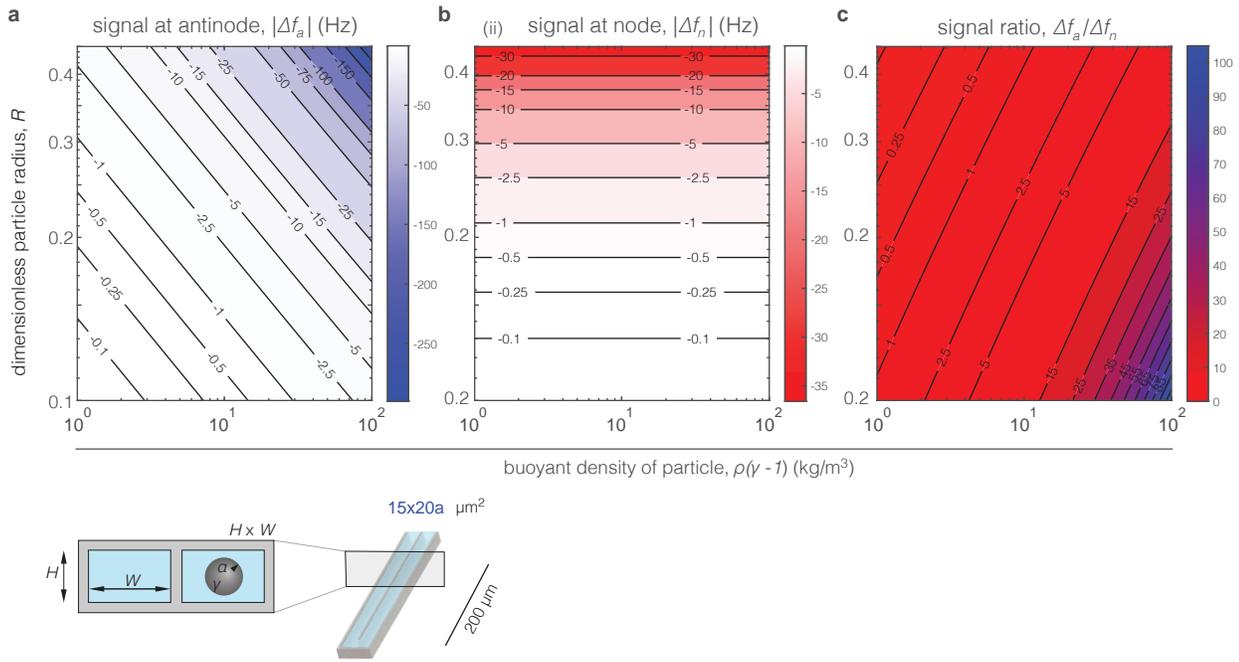

**Figure S9 | Variation of antinode and node signals across a range of particle buoyant densities and sizes relative to cantilever height.** Theoretical calculations of, **a**, $\Delta f_a$, **b**, $\Delta f_n$ and, **c**, ratio $\Delta f_a/\Delta f_n$ vs $R$ for the 15x20a cantilever (bottom schematic) with the highest, Reynolds number, $\beta$, tested experimentally (Figs. 2, 4). Ratio of the particle density to the fluid density is $\gamma = \rho_p/\rho$, thus the buoyant density is $\rho(\gamma - 1)$. For most of this range we observed that the antinode signal is greater than the node signal ($\Delta f_a/\Delta f_n > 1$).

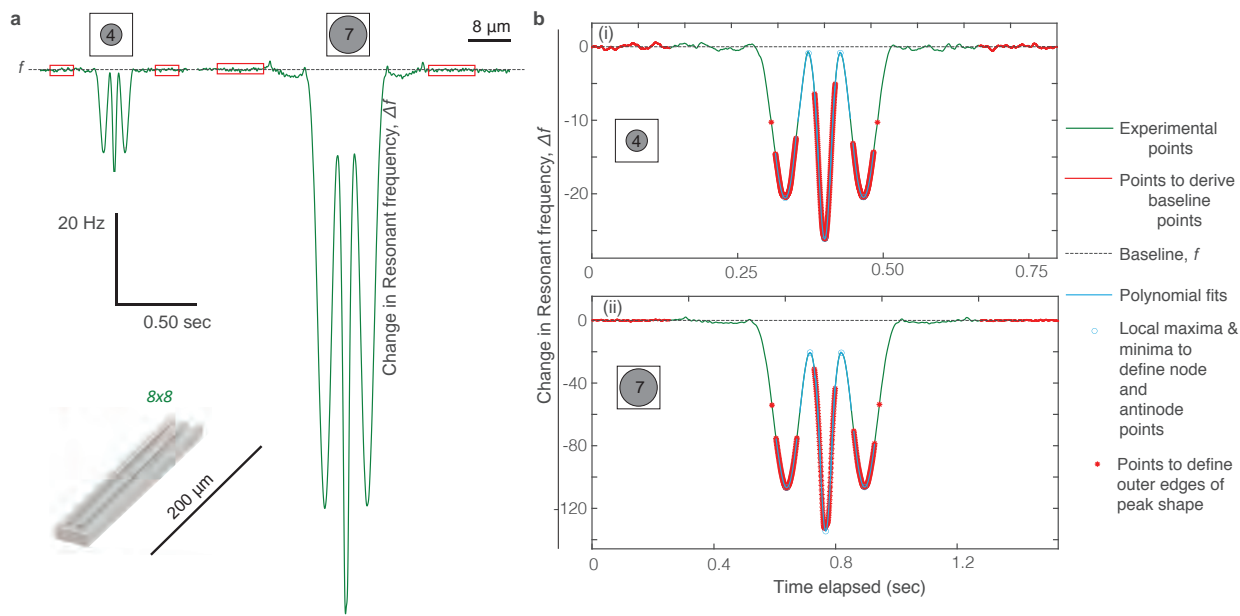

**Figure S10 | Sensitivity of the calculation of resonant frequency changes to the determination of the baseline frequency. a**, Two examples of experimentally measured resonant frequencies, $f$, in the 8x8 cantilever (same data and notation from Fig. 4a, notation '8x8') for particles of $d_{nom} = 4, 7$ μm. $f_0$ denotes the frequency of the baseline before and after the passage of the particle into the cantilever. The relative size of $d_{nom}$ with respect to the channel height, $H$, is shown in the schematics where gray circles represent the particle and black squares the cross-sectional area of the flow channel. The red rectangles define the regions before and after the passage of the particle in the cantilever, where a linear fit is done to establish the baseline frequency $f_0$. In the case of $d_{nom} = 7$ μm, where the particle size is comparable to $H$, the particle perturbs the baseline frequency before entering the cantilever, introducing an error into the calculation of $f_0$. In addition, the theory is derived in the limit $R \ll 1$, adding additional error to the interpretation of these measurements. **b**, Calculation of resonant frequency changes at the node and antinode, $\Delta f_n$ and $\Delta f_a$ respectively, from the frequency change, $\Delta f$, for (i) $d_{nom} = 4$ μm and (ii) $d_{nom} = 7$ μm. The user-defined extraction of $f_0$ introduces an uncertainty in the calculation of $\Delta f$, which affects $\Delta f_n$ more than $\Delta f_a$ due to the fact that, generally, the former has lower magnitude than the latter (Fig. 4a, S11).

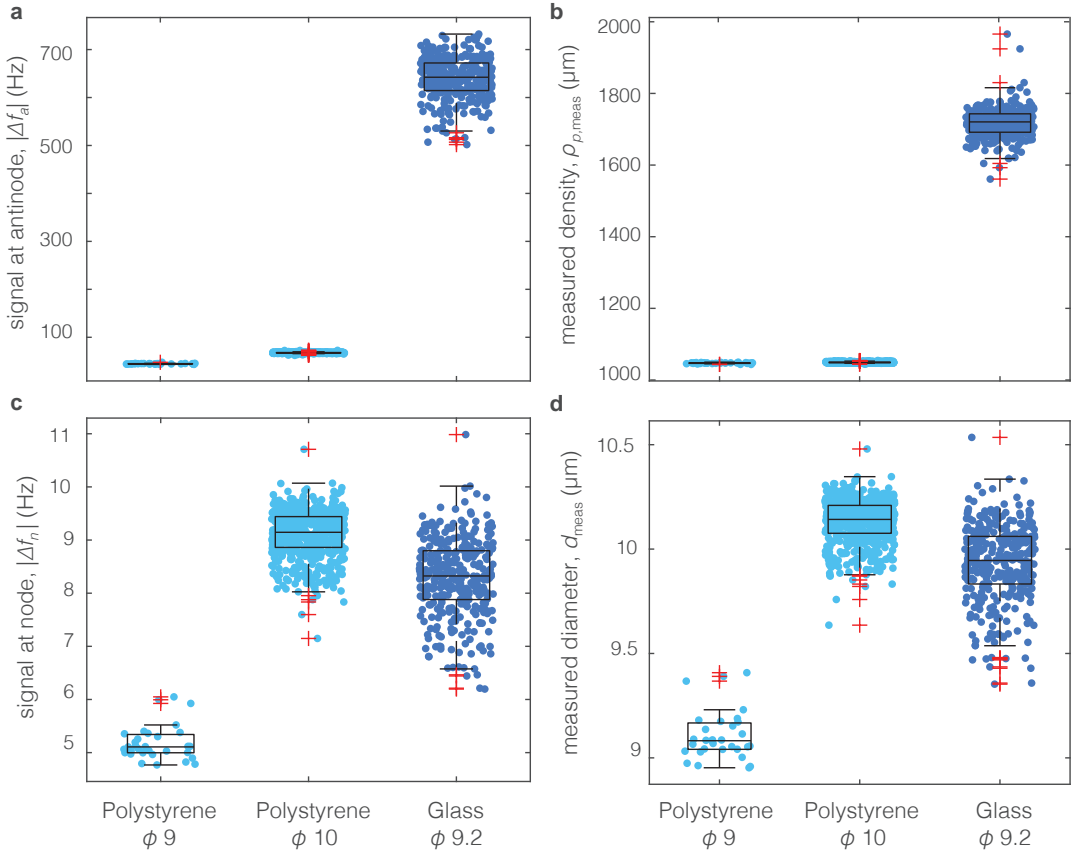

**Figure S11 | Using the node signal to measure the density of particles of different composition. a,c** Measurements of the signals, $\Delta f_n$ and $\Delta f_a$, ($\Delta f_n, \Delta f_a < 0$) at the node and antinode positions respectively (Fig. 3b) for three different particle samples in water suspensions: (1) polystyrene particles of nominal diameter $d_{nom} = 9, 10$ μm (light blue) and (2) $\rho_{p,nom} = 1,050$ kg/m³ (same data as in Fig. 4), and (3) glass microparticles of $d_{nom} = 9.2$ μm and $\rho_{p,nom} = 2,000$ kg/m³. As opposed to the polystyrene particles, the glass particles are not calibration standards and were expected to deviate from the specifications of the manufacturer, specifically, the density of these particles is not well characterized. **b,d** Measured diameter, $d_{meas}$, and density, $\rho_{meas}$, for three samples calculated using $\Delta f_n$ and $\Delta f_a$. As predicted by the theory, the glass particles have a similar node signal, $\Delta f_n$, as polystyrene particles of similar volume but an order of magnitude higher antinode signal, $\Delta f_a$, due to the corresponding difference in buoyant density $\Delta \rho_b = \rho_p - \rho_f$. The central mark indicates the mean, and the bottom and top edges of the box indicate the 25th and 75th percentiles, respectively. The whiskers extend to the most extreme data points not considered outliers (red + symbol). Experiments were performed in the 15x20a device.

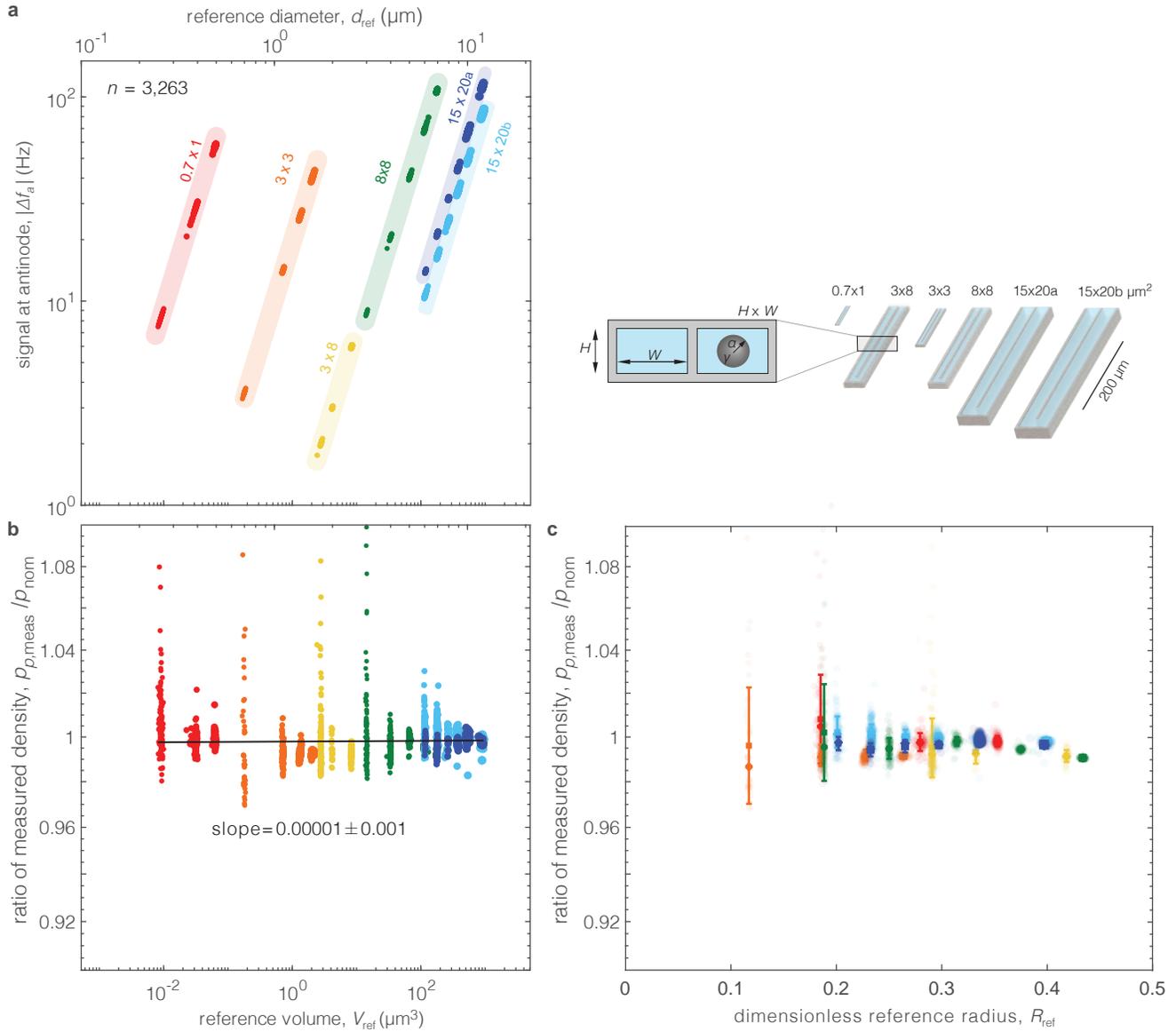

**Figure S12 | Measuring particle density from a single measurement. a**, Measurements of the frequency change signal $\Delta f_a$ at the antinode position (Fig. 3b, blue) vs reference particle diameter, $d_{ref}$, (top horizontal axis) and reference particle volume, $V_{ref}$, (bottom horizontal axis) for the same experiments shown in Fig. 4 using identical notation with colors and symbols. Note that in the present experiments, $\Delta f_a < 0$. $\Delta f_a$ is depicted without noise because it is plotted vs $d_{ref}/V_{ref}$ which are calculated directly from $\Delta f_a$ using Eq. (20) (Supplementary Note 1). **b**, Ratio of measured density $\rho_{p,meas}/\rho_{nom}$ vs $d_{ref}$ (top horizontal axis) $V_{ref}$ (bottom horizontal axis), where $\rho_{meas}$ is calculated from experimental signals of $\Delta f_n$ (Fig. 4a) and $\Delta f_a$ using the viscous enhancement factor, $a_v$ (Fig. 2), and effective mass, $m_{eff}$ (Fig. S5), using an iterative algorithm (Fig. S6). The linear regression of $V_{meas}$ is done vs $V_{ref}$ for a 99.9% confidence interval. **c**, Ratio of measured density $\rho_{p,meas}/\rho_{nom}$ vs $R_{ref} = a_{ref}/H$, where $a_{ref}$ is the reference radius of the particle; see schematic in center-right. For every dataset of given $R_{ref}$, squares denote mean, diamonds median, and error bars are standard deviation. Color notation similar to (a,b) but individual data points are shown transparent for illustration purposes.

# Supplementary Note 1: Theory

Here, we calculate the effect on the resonant frequency of a particle flowing through a suspended micro- or nanochannel resonator. The channel is assumed to have its axis aligned to the neutral axis of the resonator[18]. There are four natural length scales that arise: the velocity displacement amplitude, $\epsilon$, the particle radius, $a$, the channel height, $H$, and the cantilever length $L$. All position variables are non-dimensionalized with respect to $H$ unless otherwise specified; specifically the lengthwise position coordinate $x^*$ is scaled by $L$ (but $x$ is scaled by $H$) and the displacement function, $W(x^*)$, is scaled by $\epsilon$. All other variables are non-dimensionalized with respect to the following scales; time, $t$, by $1/\omega$ where $\omega$ is the angular frequency of the resonator (in absence of the suspended particle), velocities by $\epsilon\omega$, and pressure by $\epsilon\omega\mu/H$ where $\mu$ is the shear viscosity of the fluid; this specifies the force and torque scales to be $\epsilon\omega\mu H$ and $\epsilon\omega\mu H^2$ respectively. All dependent variables have the exponential time dependence $e^{-it}$, e.g., $\sigma = \bar{\sigma}e^{-it}$ where the real part of the expression gives the true (as measured) quantities; the superfluous overline is subsequently dropped from the analysis. A Taylor expansion of the base flow (i.e., in absence of the particle) around position $x_0$, gives[18],

$$\{\boldsymbol{u}^{(b)}, p^{(b)}\} = i\left\{-\left[\frac{H}{L}\frac{dW}{dx^*}\bigg|_{x_0^*}\left[\left(z - \frac{\sinh \lambda z}{\sinh\frac{\lambda}{2}}\right)\hat{\boldsymbol{x}} + (x - x_0)\hat{\boldsymbol{z}}\right] + W(x_0^*)\hat{\boldsymbol{z}}\right], \lambda^2 z\left(\frac{dW}{dx^*}\bigg|_{x_0^*}(x - x_0) + W(x_0^*)\right)\right\}, \quad (2)$$

where $i$ is the imaginary unit, and $\lambda = (1-i)\sqrt{\beta/2}$ where $\beta = \rho H^2 \omega/\mu$ is the oscillatory Reynolds number and $\rho$ is the fluid density.

Consider a rigid and spherical particle with radius $a$, and density $\rho_p$, located at position $(x_0, z_0)$. This specifies two more non-dimensional parameters, a density ratio $\gamma = \rho_p/\rho$ and a scaled radius, $R = a/H$ and $R^* = a/L$; throughout we assume that $R \ll 1$. This assumption allows us to linearize the flow field over the length scale of the particle, which, by using conservation of linear and angular momentum of the particle, specify its rigid body dynamics as,

$$\boldsymbol{u}_p = U_x\hat{\boldsymbol{x}} + U_z\hat{\boldsymbol{z}} + \Omega_y\hat{\boldsymbol{y}} \times \boldsymbol{r}, \quad (3)$$

where $\hat{\boldsymbol{x}}$, $\hat{\boldsymbol{y}}$ and $\hat{\boldsymbol{z}}$ are the cartesian basis vectors, $\boldsymbol{r}$ is the position vector and

$$U_x = -iR^*\frac{dW}{dx^*}\bigg|_{x_0^*}\left(z - \frac{\sinh \lambda z_0}{\sinh \lambda/2}\right)\frac{1 + R\lambda + R^2\lambda^2/3}{1 + R\lambda + (2\gamma + 1)R^2\lambda^2/9}, \quad (4)$$

$$U_z = -iW(x_0^*)\frac{1 + R\lambda + R^2\lambda^2/3}{1 + R\lambda + (2\gamma + 1)R^2\lambda^2/9}, \quad (5)$$

$$\Omega_y = iR^*\frac{dW}{dx^*}\bigg|_{x_0^*}\frac{\lambda \cosh \lambda z_0}{2 \sinh \lambda/2}\frac{1 + R\lambda + 6R^2\lambda^2/15 + R^3\lambda^3/15}{1 + R\lambda + (\gamma + 5)R^2\lambda^2/15 + \gamma R^3\lambda^3/15}. \quad (6)$$

As the particle performs the motion $\boldsymbol{u}_p$, it generates a disturbance flow due to both its own motion, and the action of the base flow on the particle surface, via the no-slip and no-penetration boundary conditions. This disturbance flow is obtained using fundamental solutions[30], and is,

$$\{\boldsymbol{u}^{(d)}, p^{(d)}\} = \left[U_x + iR^*\frac{dW}{dx^*}\bigg|_{x_0^*}\left(z - \frac{\sinh \lambda z}{\sinh \lambda/2}\right)\right]\{\boldsymbol{u}_x^{(T)}, p_x^{(T)}\} + [U_z + iW(x_0^*)]\{\boldsymbol{u}_z^{(T)}, p_z^{(T)}\} + \left[\Omega_y - iR^*\frac{dW}{dx^*}\bigg|_{x_n^*}\frac{\lambda \cosh \lambda z_0}{2 \sinh \lambda/2}\right]\{\boldsymbol{u}_y^{(R)}, 0\}$$
$$+ iR^*\frac{dW}{dx^*}\bigg|_{x_0^*}\left(1 - \frac{\lambda \cosh \lambda z_0}{2 \sinh \lambda/2}\right)\{\boldsymbol{u}^{(E)}, p^{(E)}\}. \quad (7)$$

The required fundamental solutions are given by

$$\{\boldsymbol{u}_i^{(T)}, p_i^{(T)}\} = \left(\frac{1}{4}(3 + 3R\lambda + R^2\lambda^2) - \frac{3 + 3R\lambda + R^2\lambda^2 - 3e^{R\lambda}}{4R^2\lambda^2}\right)\{u_i^{(S)}, p_i^{(S)}\}, \quad i \in \{x, z\}, \quad (8)$$

$$\{\boldsymbol{u}_x^{(S)}, p_x^{(S)}\} = \left\{A(r\lambda)\frac{\hat{\boldsymbol{x}}}{r} + B(r\lambda)\frac{x\boldsymbol{r}}{r^3}\right\} \qquad \{\boldsymbol{u}_z^{(S)}, p_z^{(S)}\} = \left\{A(r\lambda)\frac{\hat{\boldsymbol{z}}}{r} + B(r\lambda)\frac{z\boldsymbol{r}}{r^3}\right\}, \quad (9)$$

$$A(r\lambda) = 2e^{-r\lambda}\left(1 + \frac{1}{r\lambda} + \frac{1}{r^2\lambda^2}\right) - \frac{2}{r^2\lambda^2}, B(r\lambda) = -2e^{-r\lambda}\left(1 + \frac{3}{r\lambda} + \frac{3}{r^2\lambda^2}\right) + \frac{6}{r^2\lambda^2}, \quad (10)$$

$$\boldsymbol{u}^{(R)} = \frac{e^{R\lambda}}{2(1 + R\lambda)}\left(\partial_z \boldsymbol{u}_x^{(S)} - \partial_x \boldsymbol{u}_z^{(S)}\right), \quad (11)$$

$$\{\boldsymbol{u}^{(E)}, p^{(E)}\} = \frac{1}{2}\left(-\frac{15 + 15R\lambda + 6R^2\lambda^2 + R^3\lambda^3}{9(1 + R\lambda)} + \frac{15 + 15R\lambda + 6R^2\lambda^2 + R^3\lambda^3 - 15e^{R\lambda}}{9R^2\lambda^2(1 + R\lambda)}\nabla^2\right)\left(\partial_z\{u_x^{(S)}, p_x^{(S)}\} + \partial_x\{u_z^{(S)}, p_z^{(S)}\}\right), \quad (12)$$

where $r$ is the radial spherical polar coordinate. The complete flow field within the SMR/SNR (under the assumption $R \ll 1$) is specified by $\boldsymbol{u} = \boldsymbol{u}^{(b)} + \boldsymbol{u}^{(d)}$. The disturbance flow therefore alters the force and torque exerted by the fluid on the cantilever; the change to the force and torque are calculated by,

$$F^{(d)} = \hat{\boldsymbol{z}} \cdot \int_{-\infty}^{\infty}\int_{-\infty}^{\infty}\left(\boldsymbol{\sigma}^{(d)} \cdot \boldsymbol{n}\right)\big|_{z=z_0-1/2} + \left(\boldsymbol{\sigma}^{(d)} \cdot \boldsymbol{n}\right)\big|_{z=-z_0-1/2}\, dxdy, \quad (13)$$

$$M^{(d)} = \hat{\boldsymbol{y}} \cdot \int_{-\infty}^{\infty}\int_{-\infty}^{\infty}\boldsymbol{r} \times \left(\boldsymbol{\sigma}^{(d)} \cdot \boldsymbol{n}\right)\big|_{z=z_0-1/2} + \boldsymbol{r} \times \left(\boldsymbol{\sigma}^{(d)} \cdot \boldsymbol{n}\right)\big|_{z=-z_0-1/2}\, dxdy, \quad (14)$$

where $\boldsymbol{\sigma}^{(d)} \equiv -p^{(d)}\boldsymbol{I} + \nabla\boldsymbol{u}^{(d)} + \left(\nabla\boldsymbol{u}^{(d)}\right)^T$ is the usual stress tensor for an incompressible Newtonian fluid. The integration in Eq. (13) can be performed analytically and gives,

$$F^{(d)} = \frac{4\pi}{3}R^3\lambda^2(\gamma - 1)iW(x_0^*)\frac{1 + R\lambda + R^2\lambda^2/3}{1 + R\lambda + (2\gamma + 1)R^2\lambda^2/9}, \quad (15)$$

which is independent of the particle's height (i.e., $z$ position) in the channel.

Unlike the force calculation above, the torque must be integrated numerically, however analytic expressions are obtained for small and large Reynolds numbers and are,

$$M^{(d)} = -\frac{4\pi}{3} R^3 iR^* \frac{dW}{dx^*}\bigg|_{x_0^*} \begin{cases} \frac{2}{5}(\gamma - 1) + 0.2\lambda^2 & \beta \ll 1 \\ \frac{2}{3}\frac{15 + 15R\lambda + 6R^2\lambda^2 + R^3\lambda^3}{R^2\lambda^2(1 + R\lambda)} & \beta \gg 1 \end{cases}. \quad (16)$$

Note that there is also a force exerted on the cantilever in the $x$-direction which has been neglected here because it produces a frequency shift response that is one order of magnitude smaller in the cantilever displacement amplitude than the frequency shift produced by $F^{(d)}$ and $M^{(d)}$.

To calculate the effect on the resonant frequency due to the suspended particle, we examine the cantilever using Euler-Bernoulli beam theory. Under the assumption $H \ll L$, the applied force and torque act at a point along the length of the cantilever. The governing equations for the cantilever vibrations in the absence of a suspended particle, $W_0$, and in the presence of a suspended particle, $W_1$, are given by,

$$\frac{d^4 W_i}{dx^{*4}} - \frac{L^3 m_{\text{eff}}}{EI} \omega_i^2 W_i = 0, \quad i \in \{0,1\} \quad (17a,b)$$

where $m_{\text{eff}}$ is the effective mass of the resonator, $E$ its Young's modulus, $I$ its second moment of inertia and $\omega_0$ and $\omega_1$ its resonant frequency in the absence of and presence of the suspended particle respectively. The usual clamped and free boundary conditions are applied,

$$W_i(0) = 0, \quad \frac{dW_i}{dx^*}\bigg|_{x^*=0} = 0, \quad \frac{d^2 W_i}{dx^{*2}}\bigg|_{x^*=1} = 0, \quad \frac{d^2 W_i}{dx^{*2}}\bigg|_{x^*=1} = 0, \quad i \in \{0,1\}. \quad (18a,b)$$

Application of the point force and torque due to the suspended particle gives the additional boundary conditions,

$$\frac{EI}{L^2}\frac{d^2 W_1}{dx^{*2}}\bigg|_{x^*=x_0^{*-}} = -\mu\omega H^2 M^{(d)}, \quad \frac{EI}{L^3}\frac{d^3 W_1}{dx^{*3}}\bigg|_{x^*=x_0^{*-}} = -\mu\omega H F^{(d)}, \quad \frac{d^2 W_1}{dx^{*2}}\bigg|_{x^*=x_0^{*+}} = 0, \quad \frac{d^3 W_1}{dx^{*3}}\bigg|_{x^*=x_0^{*+}} = 0, \quad (19)$$

where the $-$ and $+$ superscripts refer to approach from the left and right side of the particle position, $x_0^*$ respectively. Multiplying Eq. (17a) by $W_1$, Eq. (17b) by $W_0$, taking the difference of the resulting expressions, and then integrating along the length of the cantilever leads to,

$$\Delta f = -\frac{f}{2m_{\text{eff}}}\left(m_b \, \alpha_b \, W_0^2(x_0^*) + \rho V \, \alpha_d \left(\frac{dW}{dx^*}\bigg|_{x_0^*}\right)^2\right), \quad (20)$$

where $\Delta f = (\omega_1 - \omega_0)/2\pi$ and $m_b$ and $V$ are the buoyant mass and volume of the particle, respectively. The mass discrepancy and viscous enhancement factors are given respectively by

$$\alpha_m = \Re\left[\frac{1 + R\lambda + R^2\lambda^2/3}{1 + R\lambda + (2\gamma + 1)R^2\lambda^2/9}\right], \quad (21)$$

$$\alpha_v = \Re\left[M^{(d)}\bigg/\left(-iR^2\lambda^2\frac{4\pi}{3}R^*\frac{dW}{dx^*}\bigg|_{x_0^*}\right)\right], \quad (22)$$

where $\Re$ specifies the real part of the expression. In general, $\alpha_v$ requires numeric evaluation and depends on the four parameters $R, \beta, \gamma$ and $z$. In the limits of small and large Reynolds number, $\alpha_v$ reduces to

$$\alpha_v = \begin{cases} \frac{3}{5}(\gamma - 1) + \frac{1}{20R^2}(1 - 12z^2)\sqrt{\beta} & \beta \ll 1 \\ \Re\left[\frac{15 + 15R\lambda + 6R^2\lambda^2 + R^3\lambda^3}{R^2\lambda^2(1 + R\lambda)}\right] & \beta \gg 1 \end{cases}; \quad (23)$$

Eq. (23) is used to plot the asymptotic limits in Fig. 2.

## *Influence of z-position on the viscous enhancement factor*

The disturbance flow produced by the particle (and hence also the viscous enhancement factor) depends on the gradient of the base flow at the particle's location; the gradient of the base flow specifies the rotational and extensional flow components used in the analysis presented above. If the gradient of the base flow varies across the channel height, then, in general, the viscous enhancement factor also becomes dependent on the height of the particle within the channel (Fig. S4). There are two key limits where the viscous enhancement factor becomes independent of $z$-position.

First, when $\beta \gg 1$, the base flow reduces to a purely extensional flow whose gradient is constant with respect to $z$; this can be seen by examining the $\beta \gg 1$ limit of Eq. (2). Because the gradient of the base flow is constant with respect to $z$, the viscous enhancement factor, and therefore also the frequency shift, is independent of $z$; see the $\beta \gg 1$ limit of Fig. S4 and Eq. (23).

Second, when $\beta \ll 1$, the primary component of the base flow reduces to a purely rotational disturbance flow whose gradient is, again, constant with respect to $z$. If a non-neutrally buoyant particle ($\gamma \neq 1$) is suspended in this flow, then the dynamics are dominated by the primary flow (the secondary flow may be neglected here). Because the gradient of the base flow is constant with respect to $z$, so is the viscous enhancement factor and the frequency shift; see the $\beta \ll 1$ limits of Fig. S4 and Eq. (23), for particles where $\gamma \neq 1$.

In general, when the particle moves off axis there is also a contribution to the frequency shift due to its linear velocity in the $x$-direction. As discussed above, this contribution is neglected because it is order of magnitude smaller (in terms of the cantilevers displacement amplitude) than the frequency shifts described above.

# Supplementary Note 2: Sources of experimental error

In our experiments, three main sources of error exist.

First, determination of $V_{\text{meas}}$ from $\Delta f_n$ involves a non-linear error propagation process, converting (Gaussian) frequency noise into a systematic error (Fig. 4b). Notably, the signal-to-noise ratio for $\Delta f_n$ is smaller than $\Delta f_a$ (Fig. S11). Consequently, particles of radii, $R < 0.25$, exhibit a variance in the signal, $\Delta f_n$, greater than that due to their nominal polydispersity (Table S2). By performing Monte-Carlo simulations, we conclude that this enhancement is due to their lower signal-to-noise ratio (Figs. S7, S8). When combining $\Delta f_n$ with $\Delta f_a$ to calculate density, decreasing $R$ decreases precision while not affecting accuracy (Fig. S12c). Accuracy is not correlated with $R$, indicating that there is no systematic error in applying the theory.

Second, the derived theory does not account for the bounding walls of the channel. Therefore, we expect accuracy to decrease as the particle diameter approaches the channel height ($R \to 0.5$). This error depends on the ratio of the particle radius to both the viscous penetration depth, and the channel height. Even so, it is expected to be small because the disturbance flow decays rapidly from the particle, with a power-law between 2 and 4 in the radial distance (depending on $\beta R^2$). We experimentally observed that accuracy decreases slightly for $R > 0.4$, but is still greater than 99% (Fig. S12c). Notably, we observed that for increasing $R$ there is an additional uncertainty in defining the baseline frequency, $f_0$ (Fig. S10).

Third, analysis of experimental data implicitly assumes the particle is at the channel center ($z = 0$). This also leads to error because the measured signal $\Delta f_n$, depends on the particle $z$-position; albeit at a diminishing degree when $\beta \gtrsim 100$ for the particle size range, $R > 0.20$, used here (Fig. S4). The uncertainty due to $z$-position is likely addressable by actuating the cantilevers at multiple modes simultaneously,[21] which would enable determination of the particle $z$-position.

# Supplementary Note 3: Information about movies

Movie 1: Base, disturbance and total flow fields at the node of an SMR at low Reynolds number ($\beta \ll 1$) with a heavy particle ($\gamma \gg 1$) suspended at the center of the channel; the amplitude is exaggerated for clarity. The base and disturbance flow fields correspond to Fig. 1b-i and c-i respectively. The complete flow is the actual flow within the channel and given by the sum of the base and disturbance flows.

Movie 2: Base secondary, disturbance and total flow fields at the node of an SMR at low Reynolds number ($\beta \ll 1$) with a neutrally buoyant particle ($\gamma = 1$) suspended at the center of the channel; the amplitude is exaggerated for clarity. The base secondary flow and disturbance flow correspond to Fig. 1b-ii and c-ii respectively. The total flow is the sum of the base secondary and disturbance flows; note that the actual flow field within the SMR also contains a contribution from the primary base flow (Fig. 1b-i).

Movie 3: Base, disturbance and total flow fields at the node of an SMR at high Reynolds number ($\beta \gg 1$) with a (arbitrary density) particle suspended at the center of the channel; the amplitude is exaggerated for clarity. The base and disturbance flows correspond to Fig. 1b-iii and c-iii respectively. The complete flow is the actual flow within the channel and given by the sum of the base and disturbance flows. Note that the (small) apparent violation of the no-penetration condition at the walls is an artifact of visualizing the small amplitude solution at large amplitude.

Movie 4: Two experiments with the 8x8 (left) and 15x20b (right) devices show polystyrene particles of nominal diameter 7 $\mu$m entering the cantilevers (top) and generating a change in the recorded frequency (bottom). The signal of frequency change is different based on the channel Reynolds number, $\beta$, the baseline resonant frequency, $f$, and the characteristics of the each device (Table S1, Fig. S5). The data are from Fig. 4 with same color notation for referring to the type of device.